\documentclass[conference]{IEEEtran}
\IEEEoverridecommandlockouts
\usepackage{cite}
\usepackage{amsmath,amssymb,amsfonts}
\usepackage{algorithmic}
\usepackage{graphicx}
\usepackage{textcomp}
\usepackage{xcolor}

\usepackage{cite}
\usepackage{balance}
\usepackage{textcomp}
\ifCLASSINFOpdf
\else
\usepackage[dvips]{graphicx}
\fi

\usepackage{algorithm}      
\usepackage{algorithmic}

\usepackage{varwidth}

\begin{document}


\title{A Maneuver-based Urban Driving Dataset and Model for Cooperative Vehicle Applications}

\author{Behrad Toghi$^*$, Divas Grover$^*$, Mahdi Razzaghpour$^*$, Rajat Jain$^*$, Rodolfo Valiente$^*$, Mahdi Zaman$^*$,\\
Ghayoor Shah$^*$, Yaser P. Fallah$^*$\\
$^*$Connected and Autonomous Vehicle REsearch Lab (CAVREL),\\
University of Central Florida, Orlando, FL,\\
toghi@knights.ucf.edu
}



\maketitle

\begin{abstract}

Short-term future of automated driving can be imagined as a hybrid scenario in which both automated and human-driven vehicles co-exist in the same environment. In order to address the needs of such road configuration, many technology solutions such as vehicular communication and predictive control for automated vehicles have been introduced in the literature. Both aforementioned solutions rely on driving data of the human driver. In this work, we investigate the currently available driving datasets and introduce a real-world maneuver-based driving dataset that is collected during our urban driving data collection campaign. We also provide a model that embeds the patterns in maneuver-specific samples. Such model can be employed for classification and prediction purposes.
\end{abstract}
\section{Introduction}
Connected and automated vehicles (CAVs) have received significant attention during the last decade. Especially, the rise of artificial intelligence and sophisticated machine learning algorithms sped up the research and development of CAVs. Commercial level-4 autonomous vehicles \cite{sae:autonomouslevels} are expected to emerge in the market as early as mid 2020s which will lead to experiencing a hybrid artificial intelligence (AI)-human scenario. In such hybrid scenarios, autonomous and human-driven vehicles co-exist on and share the same road infrastructure, and most importantly, interact with each other. The aforementioned interaction translates to the concept of human-agent cooperation in mixed-autonomy scenarios, agents, i.e., autonomous vehicles, have an internal model of human behaviors \cite{mahjoub2018stochastic} and employ that to "manipulate" the behavior of human driven vehicles, creating a potential cooperation among agents and humans \cite{sadigh2018planning, sadigh2016planning}. Furthermore, by creating an understanding of human driving patterns, autonomous vehicles are able to act in a predictive and proactive fashion in order to prevent crashes and safety-critical situations.

On the other hand, with regards to connected and cooperative vehicles, agents share their situational awareness over ad-hoc vehicular networks (VANETs), taking advantage of vehicle-to-vehicle (V2V) communication technologies such as Dedicated Short-range Communication (DSRC) \cite{kenney2011dedicated, shah2019real} and Cellular Vehicle-to-everything (C-V2X) \cite{toghi2018multiple, toghi2019analysis, toghi2019spatio}. Recently, authors in \cite{yfallah:mbcsyscon, mahjoub2018driver} suggested a novel methodology for V2V communication known as the model-based communication (MBC). The main idea behind MBC is utilizing an abstracted form of the vehicles' situational awareness, i.e., an abstract model of their state, as an alternative for the current standard raw-data communication.

A vehicle's mobility patterns can be classified mainly into a dozen of short-term, or small-scale, maneuvers. Among which, one can refer to maneuvers such as U-turns, lane changes, left (and right) turns, hard-brakes, joining (and leaving) a platoon, take-over, etc. Precise detection of such maneuvers enables engineers and researchers to design robust safety and collision avoidance systems for automated vehicles. Modeling a maneuver provides us with an abstract representation of the vehicle's state which can be utilized within the context of earlier discussed MBC framework. Furthermore, in a hybrid human-agent scenario, recognizing a remote human-driven vehicle's intention to perform a maneuver will enable autonomous vehicles to predict and react accordingly \cite{mahjoub2019v2x}.

\begin{figure}[t]
\centering
\includegraphics[width=\columnwidth]{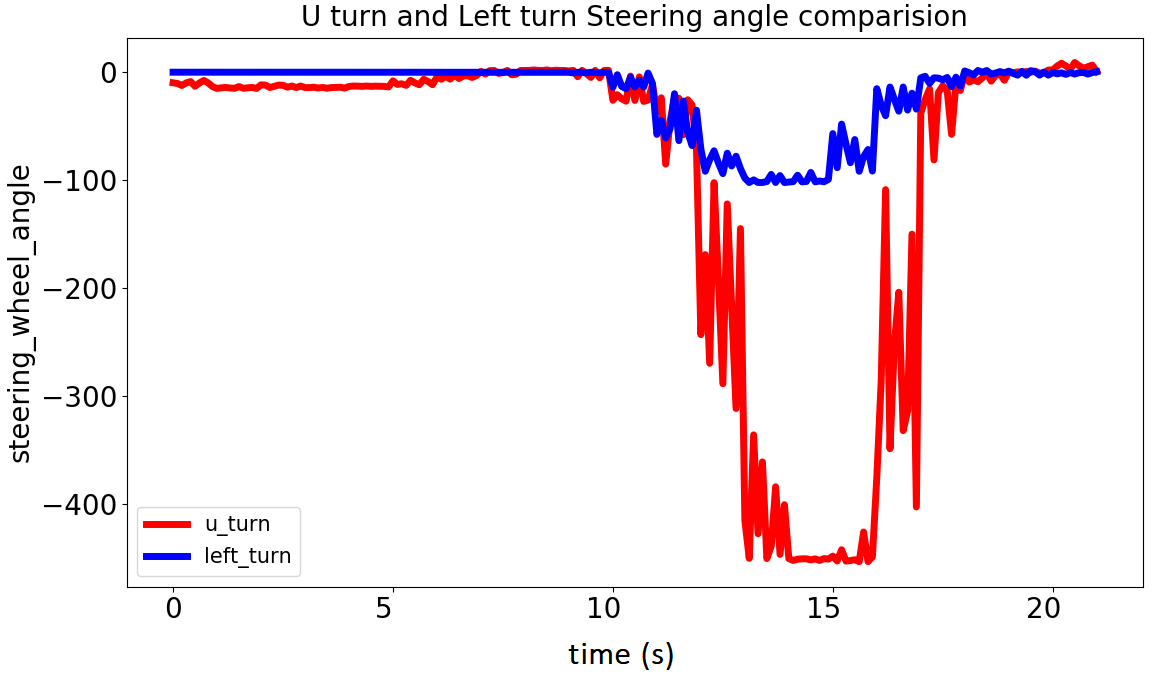}
\caption{A comparison between a pair of given u-turn and left-turn maneuvers and their pattern in the steering angle space and time-domain representation (sampled with 10 Hz rate).}
\label{fig:maneuverComparison}
\end{figure}

A wide spectrum of sensory data types is available from each vehicle's Controller Area Network (CAN) bus which can be employed to model the mobility patterns and driving maneuvers. As an instance of the most common features, one can name steering angle, engine speed, GPS coordinates and heading, and throttle position. Every specific maneuver has different class-correlation with the features and thus can be highly correlated to one and uncorrelated from the other. As an illustration, a u-turn maneuver is more significantly observable in the steering-angle/heading domain whereas a hard brake maneuver stands out in the ground speed space. Figure \ref{fig:maneuverComparison} shows the steering angle pattern in a given pair of u-turn and left-turn maneuvers.

The rest of the paper is organized as follows. Section II presents a brief review on the related work that exist in the literature. In Section III, we demonstrate the field test and data collection process and describe the dataset architecture. In Section IV, we focus on the implemented maneuver classification algorithms and present the results and analysis before concluding the paper in Section V.

\section{Related Work}
It is expected that expressing driving maneuvers using the above mentioned features will include redundant information. From an information theory point-of-view this redundancy enables us to achieve high accuracy both in classification and regression operations. A few real-world driving datasets currently exist in the literature which contain recorded \{GPS+CAN\} data from human driven vehicles in urban and highway environments. Among which, we can refer the SPMD dataset \cite{SPMD} recorded in Ann Arbor and Greater Detroit area in Michigan, as well as the 100-car near-crash dataset \cite{100car} which focuses on critical near collision scenarios and can be used for the safety related applications such as forward collision warning (FCW), and NGSIM dataset \cite{NGSIM} which is extracted from video footage of a highway and includes short-term maneuvers of vehicles in the field of view of the highway cameras.

As opposed to the work that focus on the CAN-bus or GPS data, a variety of research works are interested in the computer vision aspects of cooperative vehicles and vehicular communications. Among which, one can refer to the recent work by the authors in \cite{valiente2020dynamic} which proposes the idea of sharing Dynamic Object Maps among vehicles to be used in cooperative vehicle safety applications. Authors in \cite{valientea2020connected} study a method based on deep-learning that enables vehicles to share their situational awareness. The closest perspective to our approach is probably in \cite{valiente2019controlling} in which the authors control an automated vehicle's steering utilizing a long short-term memory (LSTM) deep network and camera view that is shared between the vehicles.

None of the mentioned works have parsed and labeled the data into specific maneuvers which adds a burden for the researchers to manually label the maneuvers in the post-process stage. A main downside of post-process data labeling is the low reliability and likelihood of false labeling which can degrade the desired regression or classification application's performance. In this work, we present a maneuver-based real-world driving dataset for the CAV applications, titled \textit{Driving Dataset for Connected and Automated Vehicles} (D$^2$CAV)\footnote{\textbf{Available online on: https://github.com/BehradToghi/D2CAV}}. The D$^2$CAV dataset contains a large set of logged CAN bus and GPS data from human-driven vehicles performing a variety of maneuvers in the Orlando metro area in Florida. We limited our interest mainly to a narrowed set of maneuvers, i.e., left (and right) turns at intersections, u-turns, hard-brakes, lane changes, and approaching intersections.

\begin{figure}[t]
\centering
\includegraphics[width=0.7\columnwidth]{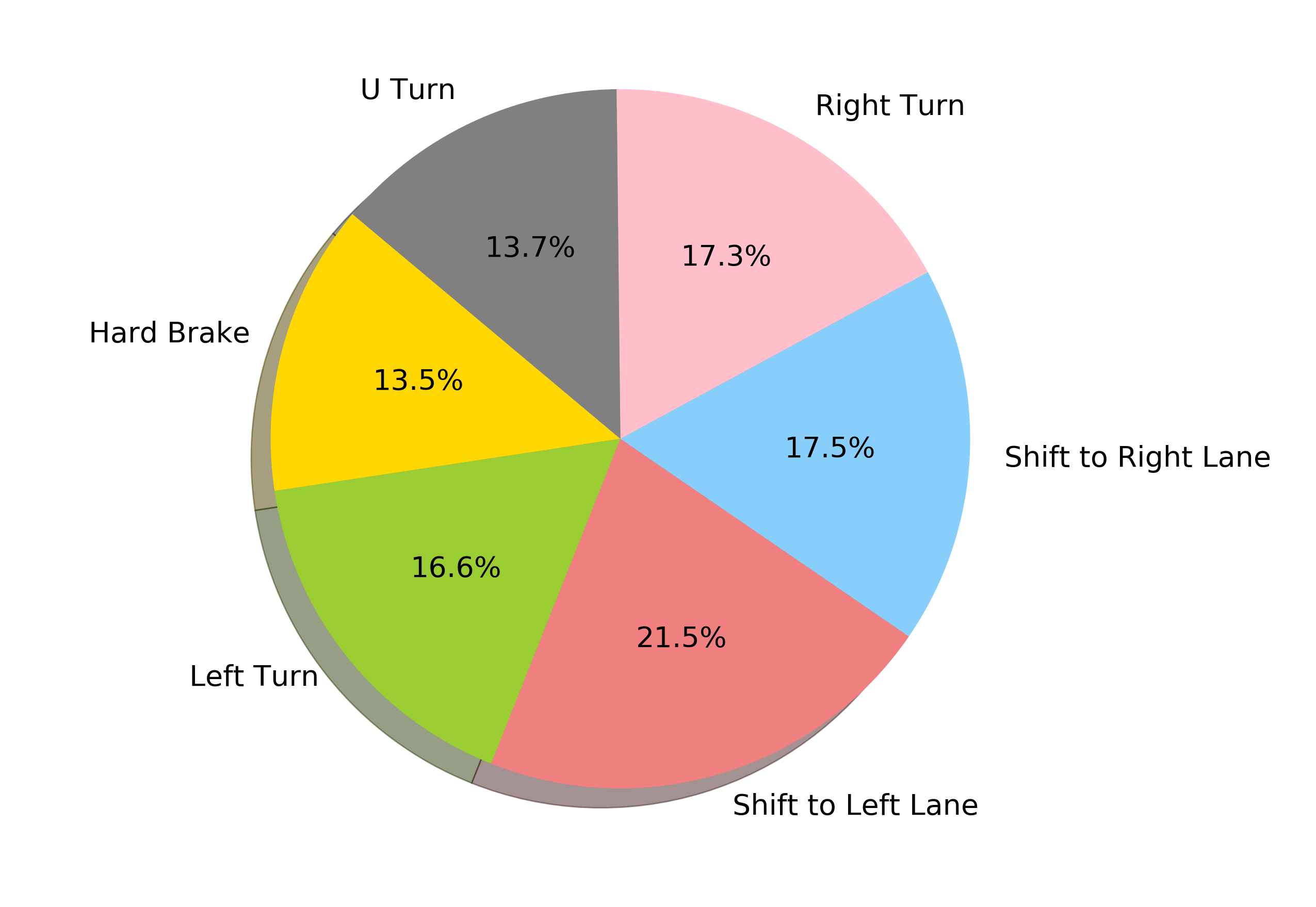}
\caption{Imbalanced dataset distribution for containing maneuvers}
\label{fig:pieChartClassDistribution}
\end{figure}

\section{In-field Data Collection Campaign}

For the purpose of data collection, we utilized the Ford OpenXC \cite{openxc} platform in addition to a Garmin Map-62s handheld unit as the logging tools. Three drivers with different driving styles (aggressive, moderate, and conservative) are asked to drive a 2018 Ford Focus equipped with electric assist steering and drive-by-wire throttle actuator. Our data collection team performed $\sim1000$ minutes of urban and highway driving around the University of Central Florida (UCF) campus in the metro Orlando area. During the field test, a co-pilot was trained and assigned to manually label the maneuvers using a custom-made logging interface, designed and implemented in Connected \& Autonomous Vehicle REsearch Lab (CAVREL).

\begin{figure*}[t!]
\centering
\includegraphics[width=0.9\textwidth]{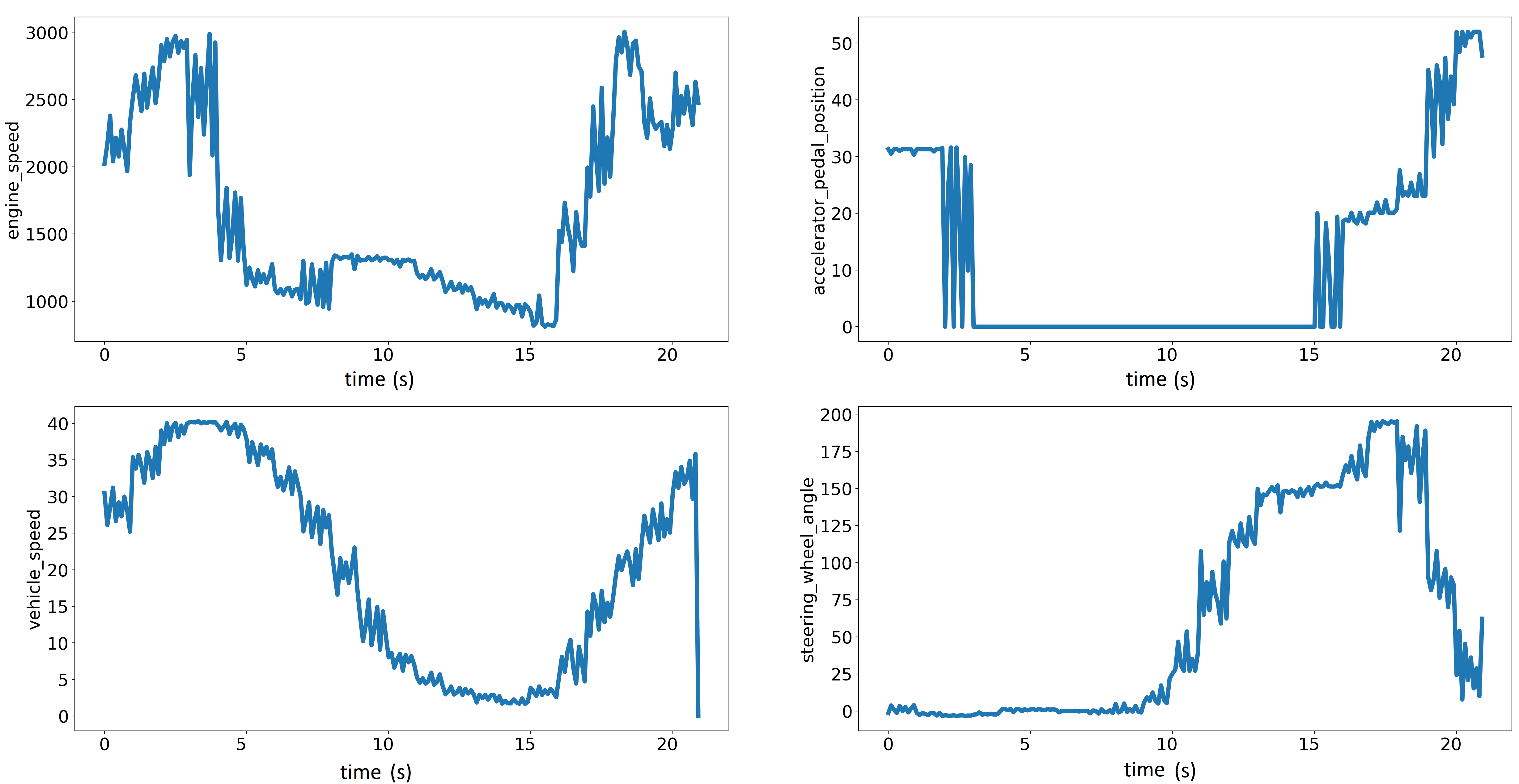}
\caption{Illustration of a given right-turn maneuver in terms of 4 features, represented in time-domain (sampled with 10 Hz rate).}
\label{fig:subplots}
\end{figure*}

The logged data fields include engine speed, total fuel consumption since restarting the vehicle, odometer, accelerator pedal position, torque at transmission, steering wheel angle, vehicle speed, and fuel level recorded by the OpenXC logger in addition to latitude, longitude, ground speed, and heading recorded by the Garmin handheld GPS device. As a matter of fact, this large number of features provide us with a set of redundant data which can potentially improve the performance of the applications implemented and trained utilizing our dataset.

\begin{figure}[b]
\centering
\includegraphics[width=0.8\columnwidth]{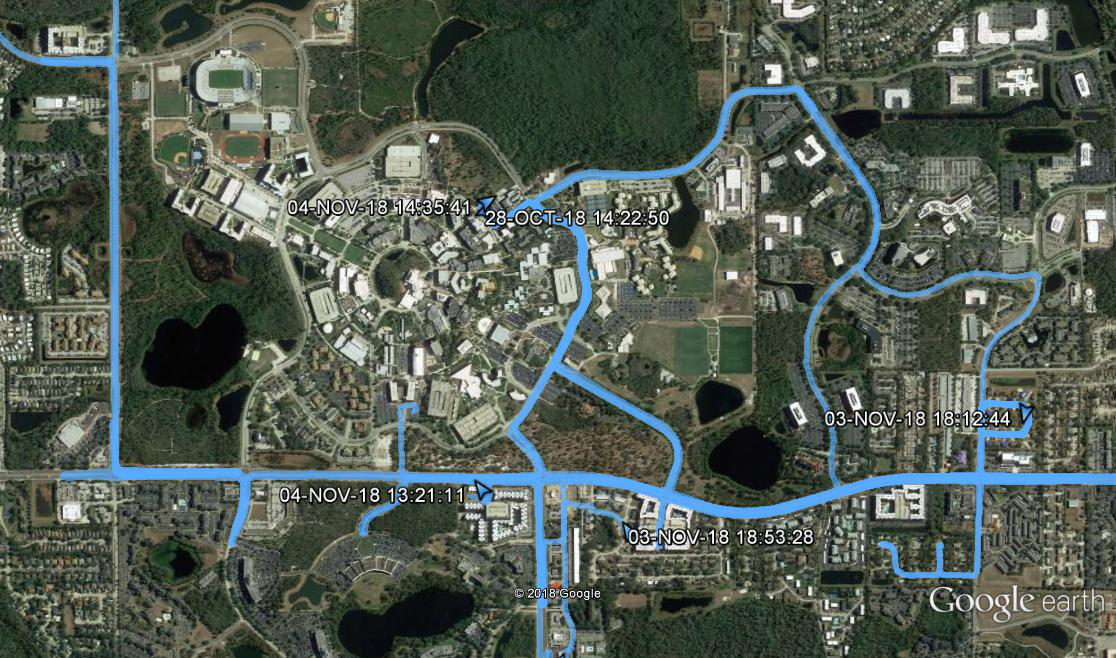}
\caption{A sample view of the driving path during the data collection campaign in the UCF campus (map courtesy of Google Earth\texttrademark)}
\label{fig:googleMap}
\end{figure}

Our setup includes the OpenXC logger connected to the vehicle's OBD II connector, the handheld Garmin GPS mounted on the windshield, and the labeling operator. Prior to performing a driving maneuver, the driver notifies the co-pilot about his intention and asks them to log the label via the logging interface. The logging interface automatically acquires the timestamp and records the label to be used in the post-processing stage. Different maneuvers can take different lengths of time, as an example, a u-turn is usually a longer maneuver (in the time domain) compared to a hard-brake maneuver. Hence, we set a $\pm10 s$ window for each maneuver and parse the trip data into smaller sub-trips, each of which contain an isolated driving maneuver. The data is organized with a straight-forward arrangement as follows. Each sub-directory contains the ".csv" file of the joint GPS+CAN data of the maneuver as well as the time-domain plots of the features and a schematic of the sub-trip, i.e., the geographic representation of the maneuver.

Figure \ref{fig:googleMap} shows the recorded driving path for the full dataset illustrated on top of the Google Earth\texttrademark , which contains a wide variety of driving conditions and maneuvers. Considering the fact that we utilized two different data sources, i.e., Vehicle CAN bus and GPS device, leads us to a challenge in the data collection campaign. These data sources not only have different data rate but also are not synchronized in the time-domain. In fact, the GPS logs have an average update rate in the order of 1 Hz, where the CAN bus data is mostly consistent on 10 Hz rate. Thus, aggregating the CAN and GPS data logs is not straight-forward and in order to address this issue, we interpolated, i.e., up-sampled the GPS logs and synthetically created timestamps to match with the CAN logs. Trying different interpolation methods showed us the cubic interpolation provides us with a more realistic vehicle mobility behavior. 

To summarize the discussion on dataset introduction, we have demonstrated the unbalanced class distribution in Figure \ref{fig:pieChartClassDistribution}. It should be noted that higher precision is expected to be achieved in applying either regression or classification methods on some maneuvers such as u-turns in comparison to the less visible (in the recorded data) maneuvers such as lane changes. This matter will be more elaborated in the next section. Each sample scenario contains the time series of the aforementioned logged features, e.g., latitude, longitude, steering angle, etc. Figure \ref{fig:subplots} shows an example sub-trip data of an arbitrary right-turn maneuver.

\subsection{OpenXC Platform}
The Research and Innovation Center in Ford Motor Company have developed a new logging interface compatible with all new-model Ford vehicles in order to support the research requirements in the academia and industry. The project is named the OpenXC \cite{openxc} platform and includes an OBD II CAN bus logger and data analysis API. The API can be run on both Ubuntu machines as well as the Android cell phones which provided us with more flexibility in the data collection process.

\section{Classification Methodology}
\begin{figure}[t]
\centering
\includegraphics[width=\columnwidth]{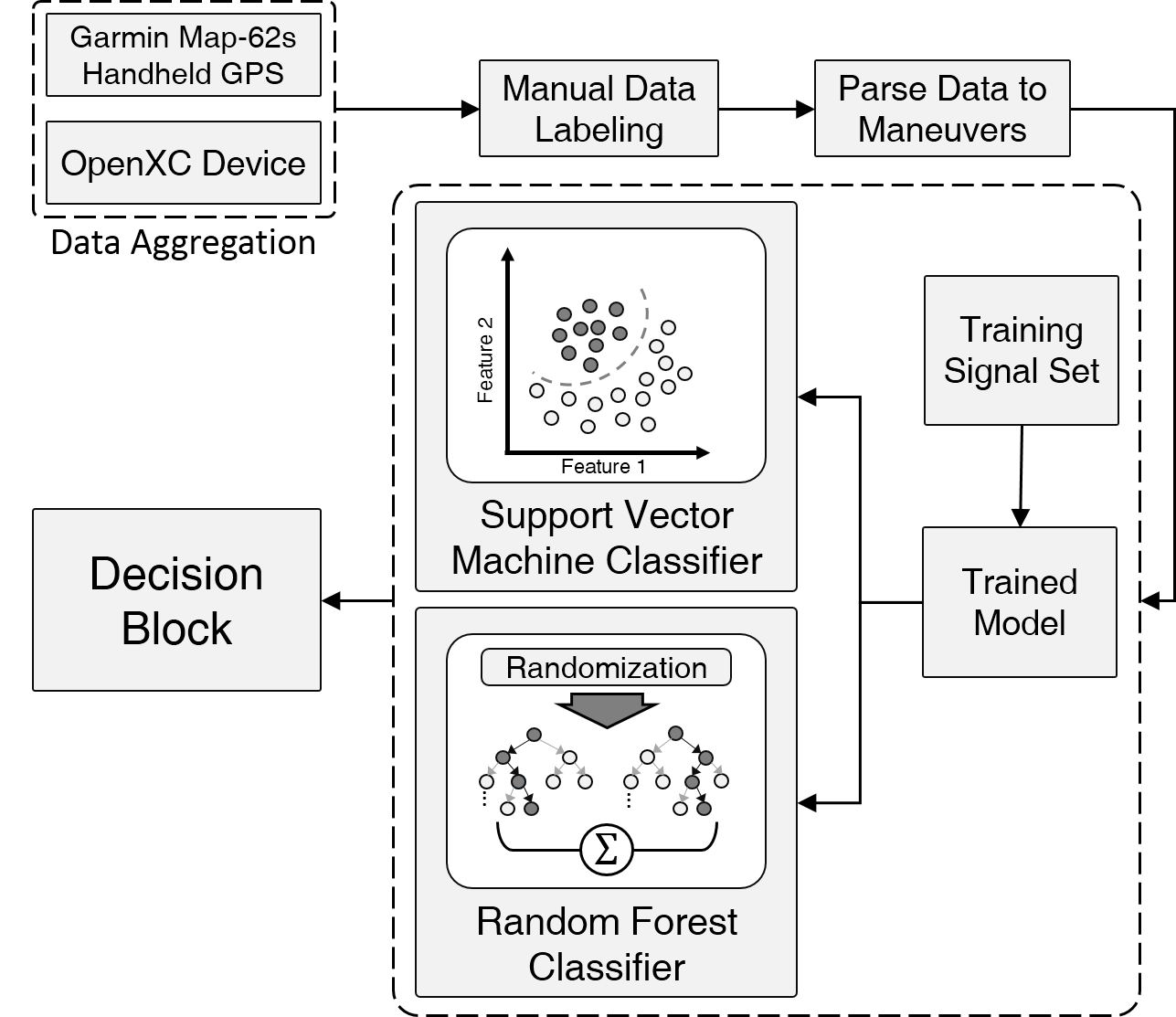}
\caption{Maneuver classification system architecture: Random Forest classifier and Support Vector Machine}
\label{fig:classifier}
\end{figure}

As it is mentioned earlier in the text, the dataset can be utilized for both prediction (regression) and classification purposes. However, in this work we focus on the latter and apply two common classification methods on the dataset and measure their performance for different maneuvers. We choose Random Forest and Support Vector Machine classifiers as the candidates and compare their performance to make a decision as the final classifier to be used as the decision block. Figure \ref{fig:classifier} shows an overview of the system architecture of our approach.

\subsection{Random Forest Classifier}
A Random Forest Classifier (RFC)~\cite{breiman2001random} is a tree based classifier which combines multiple weak learners, decision trees, to produce a strong learner, so it falls into the family of ensemble learning algorithms. We choose RFC as one of the candidates for classification as it is well-known to be robust with regards to noise, bias and over-fitting. Random Forest belongs to a class of perfectly high performing and unambiguous decision maker, therefore, it can significantly tackle on the higher-order data set which are highly-correlated as the case for the D$^2$CAV dataset.

Random Forests are trained via the bagging method. Bagging or Bootstrap Aggregating, consists of randomly sampling subsets of the training data, fitting a Decision Tree to these smaller data sets, and aggregating the predictions. This method allows several instances to be used repeatedly for the training stage given that we are sampling with replacement.

Consider the learning set represented as $\{(X_1,Y_1),\dots, (X_n,Y_n)\}$ which is n i.i.d. observation from a random vector $\mathbf(X,Y)$. Vector $X=(x_{1},\dots, x_{m})$ $X \in \mathbb{R}^{m}$ contains predictors, and $Y \in c$ where $c$ is class labels.  A classifier $T$ is a mapping from $\mathbb{R}^{m}$ to $c$. A decision tree classifier routes the input feature $x_{i}\in X$ from the root of the tree to its leaf. The final class prediction pertaining to the feature $x_i$ can be obtained at the leaf $L(T_j(x_i))$, where $T_j$ corresponds to a tree with an index $j$. 

The predicted class for a new data is calculated by majority vote of trees for that data, which results in more accuracy and stability.

\subsection{Support Vector Machine}
Support Vector Machines (SVMs) are of most simple, yet efficient, classifiers that can be applied on both linearly and non-linearly separable data. The SVM classifier enjoys a bound on the test error rate and can also employ complex non-linear kernels such as Radio Basis Functions (RBF) and exponential kernels. Therefore, we chose SVM as our second classifier candidate to be trained using the D$^2$CAV dataset.

The SVM classifier simply relies on maximizing the margin between the classifier hyper-plane and the support vectors. The well-known kernel trick can be utilized in order to apply non-linear hyper-planes. The SVM algorithm can be mathematically formulated as follows

\begin{equation} 
\begin{split}
\text{minimize:}\quad \Phi(w)=\frac{1}{2}w^Tw \\
 \text{Subject to:}\quad y_i(wx_i+b)\ge1
\end{split}
\end{equation}

where $w$ is the weight vector, $x_i$ is the input data and $y_i$ is the corresponding label. 

\begin{figure}[b]
\centering
\includegraphics[width=0.8\columnwidth]{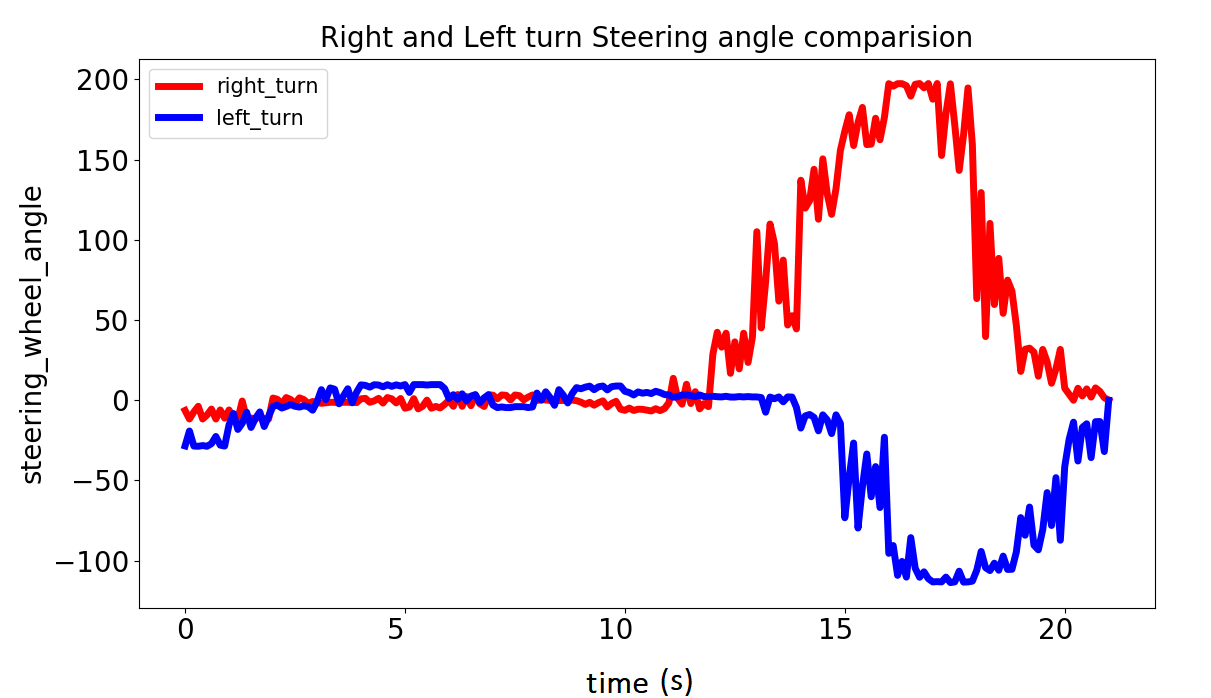}
\caption{A comparison between the left-turn and right-turn maneuvers and their time-domain (sampled with 10 Hz rate).}
\label{fig:RturnLturn}
\end{figure}

\section{Analysis \& Results}
As discussed before, we implemented two classification algorithms on our dataset and carefully measured the performance of each classifier. Before exploring the results and analysis, it is worth mentioning that one may arise the question that why the left and right turn maneuvers are being considered as two separate and independent maneuvers while they collectively can just simply be referred to as "turns". In order to address this question, we made observations on the patterns of a given left and right turn scenarios from a driving trip. The steering-angle time-series are illustrated in Figure \ref{fig:RturnLturn} for the sake of comparison and as it is obvious from the figure, left and right turn maneuvers are not exactly symmetric. The simple reason behind this is the geometry of our roads, as an example in an intersection of a left-hand-drive road, the driver needs to traverse a larger radius circle in the left-turn in comparison to the right-turn. This is also in consistency with the results shown in Figure \ref{fig:RturnLturn}.

\begin{figure}[t]
\centering
\includegraphics[width=\columnwidth]{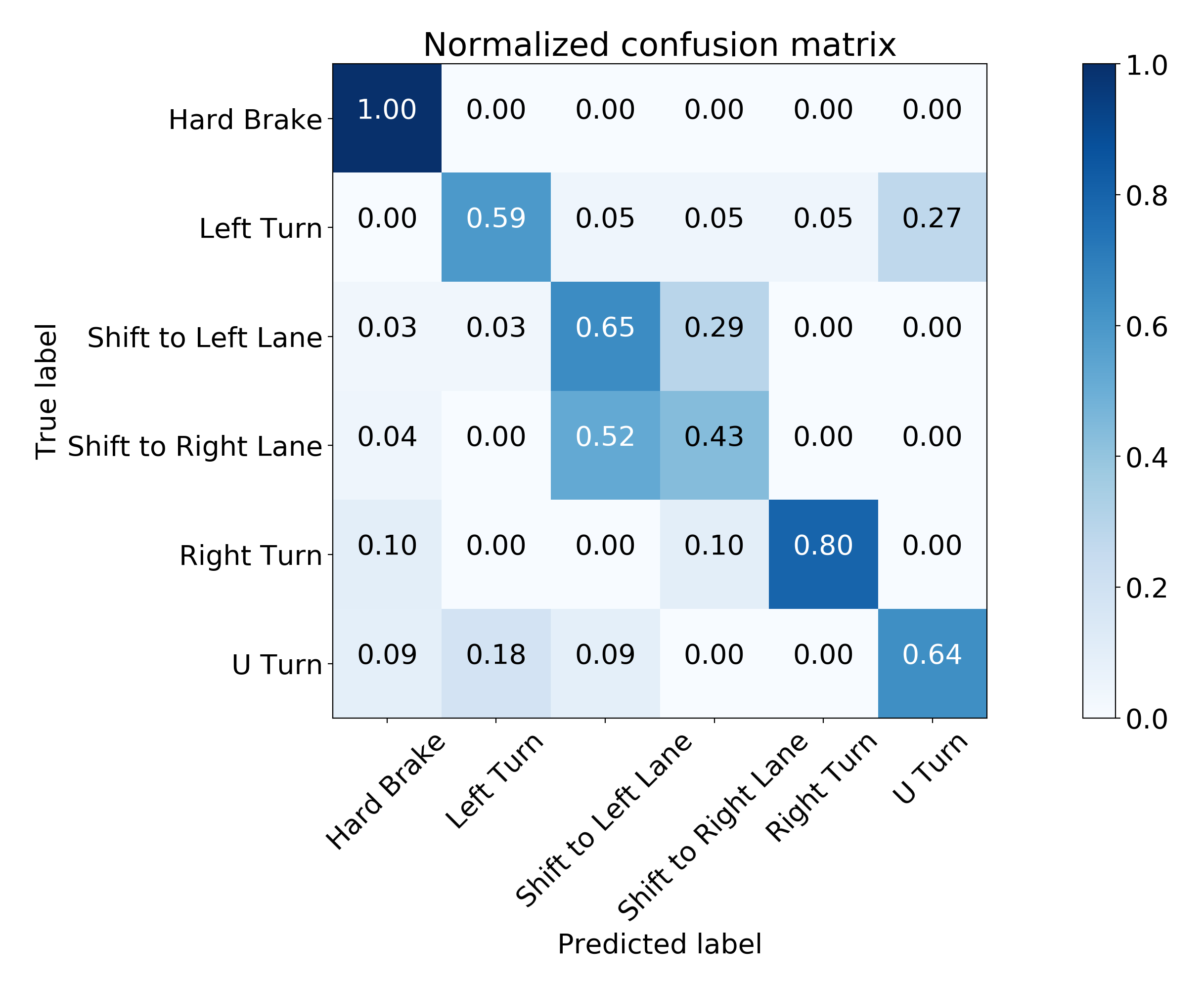}
\caption{The confusion matrix demonstrating the classification performance of the SVM classifier.}
\label{fig:confMatSVM}
\end{figure}

\begin{figure}[t]
\centering
\includegraphics[width=\columnwidth]{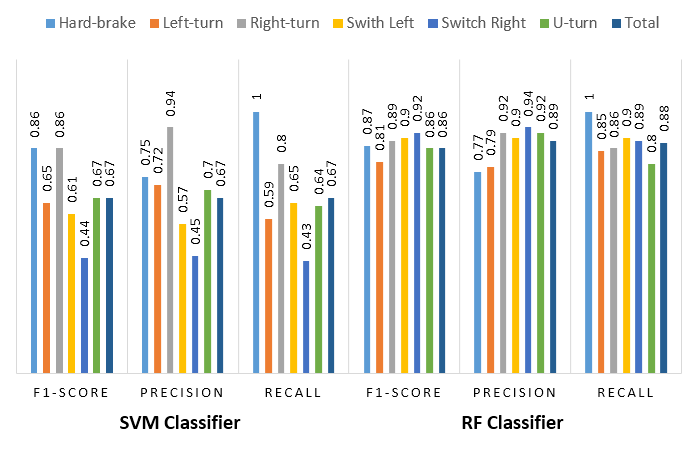}
\caption{Performance comparison of the RF and SVM classifiers for each of driving maneuvers.}
\label{fig:resultBarPlot}
\end{figure}

In order to evaluate the classification performance, we utilized three key performance indicators: F1-score, precision, and recall. Moreover, we plotted the confusion matrices to demonstrate the classification performance and errors for each of the maneuvers. Figure \ref{fig:resultBarPlot} compares performance metrics for both SVM and RF classifiers. As it is shown in the bar plots, the SVM classifier demonstrates a noticeably lower performance, almost in all metrics, compared to the RFC case. Specifically, the SVM classifier suffers in the case of left (and right) lane changes and is not able to correctly classify most of the maneuvers. This may happen due to the intrinsic similarity between two maneuvers. On the other hand, the same problem is visible in the confusion matrix shown in Figure \ref{fig:confMatSVM} where most of the misclassified data samples belong to the left and right switches. Another interesting observation can be made from the confusion matrices. Both SVM and RFC can classify all hard brake instances while there is a very plausible reasoning behind this. During the hard brake, the driver takes off their foot from the accelerator pedal and puts high pressure on the brake pedal which in turn leads to accelerator pedal position being dropped to zero alongside the engine speed and vehicle speed decreasing rapidly. This result from the confusion matrices is critical to safety applications in CAVs as a hard brake signs a near-critical case.

\begin{figure}[t]
\centering
\includegraphics[width=\columnwidth]{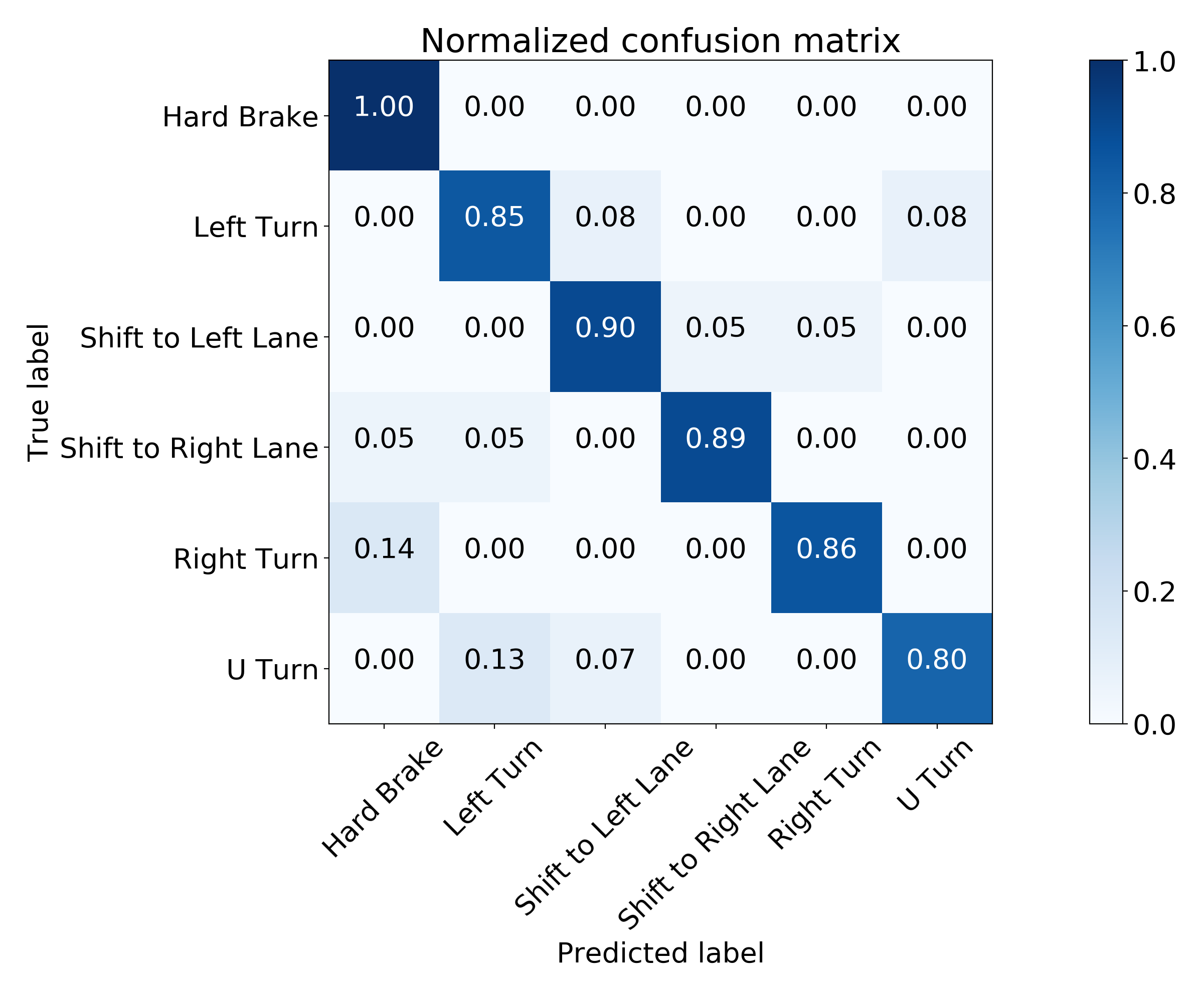}
\caption{The confusion matrix demonstrating the classification performance of the Random Forest classifier.}
\label{fig:confMatRF}
\end{figure}

The RFC performs better and demonstrates satisfactory results in terms of classification performance. The F1-score, recall, and precision metrics are shown in Figure \ref{fig:resultBarPlot} which shows that RFC is able to maintain $>80\%$ measure in all performance metrics. Moreover, the confusion matrix for the RFC case is plotted in Figure \ref{fig:confMatRF} and obviously shows a more diagonal distribution which translates to a better classification performance. An interesting and informative discussion could be analyzing the misclassified cases as some of the maneuvers have sub-maneuvers in common. As an instance, a right-turn maneuver is mostly accompanied by a break before performing the turn. Thus, if the data is not precisely parsed, we may see more misclassified data samples in this case.

\section{Concluding Remarks}

With the introduction of Connected and Automated Vehicles and consequently novel technologies for addressing the technical problems, such as the model-based communication (MBC) and predictive decision making, the need for human-driven driving datasets is arising. In this work, we investigated the currently available datasets and concluded that the literature lacks a driving dataset in which each data sample is parsed with regards to driving maneuvers, e.g., left-turn, u-turn, lane-change. We employed the Ford OpenXC in-vehicle logging platform for our data collection campaign and recorded urban driving CAN-bus and GPS data. Such maneuver-specific dataset enables the future work to benefit from the existing patterns and commonalities among driving maneuvers. Finally, two well-known classification algorithms, i.e., Support Vector Machine (SVM) and Random Forest Classifier (RFC) are implemented and trained using our dataset and their performance is evaluated on each maneuver. We discuss the results and show how such trained models can be utilized in cooperative driving applications. 

\section{Acknowledgement}
This research was supported in part by the National Science Foundation under grant number CNS-1932037.
Moreover, We would like to appreciate the support from Ford Motor Company and OpenXC \cite{openxc} research team for providing us with  OpenXC logging instrument.

\balance
\bibliography{refs.bib}{}
\bibliographystyle{unsrt}

\end{document}